# One-at-a-time: A Meta-Learning Recommender-System for Recommendation-Algorithm Selection on Micro Level

Andrew Collins, Dominika Tkaczyk and Joeran Beel

ADAPT Centre, School of Computer Science and Statistics, Trinity College Dublin, Ireland
ancollin@tcd.ie, d.tkaczyk@gmail.com,
Joeran.Beel@adaptcentre.ie

**Abstract.** The effectiveness of recommendation algorithms is typically assessed with evaluation metrics such as root mean square error, F1, or click through rates, calculated over entire datasets. The best algorithm is typically chosen based on these overall metrics. However, there is no single-best algorithm for all users, items, and contexts. Choosing a single algorithm based on overall evaluation results is not optimal. In this paper, we propose a meta-learning-based approach to recommendation, which aims to select the best algorithm for each user-item pair. We evaluate our approach using the MovieLens 100K and 1M datasets. Our approach (RMSE, 100K: 0.973; 1M: 0.908) did not outperform the single-best algorithm, SVD++ (RMSE, 100K: 0.942; 1M: 0.887). We also develop a distinction between meta-learners that operate per-instance (micro-level), per-data subset (mid-level), and per-dataset (global level). Our evaluation shows that a hypothetically perfect micro-level meta-learner would improve RMSE by 25.5% for the MovieLens 100K and 1M datasets, compared to the overall-best algorithms used.

## 1 Introduction

The 'algorithm selection problem' describes the challenge of finding the most effective algorithm for a given recommendation scenario. Some typical recommendation scenarios are news websites [3], digital libraries [4, 5], movie-streaming platforms [13]. The performance of recommender system algorithms vary in these different scenarios [3, 6, 10, 11, 15] as illustrated in **Fig. 1**. Performance variation occurs for many reasons, for example, the effectiveness of collaborative filtering algorithms changes depending on the number of ratings available from users [10]. Algorithms also perform differently depending on the demographic characteristics of users [6][11], depending on the time of the day that recommendations are delivered, the number of requested recommendations, and many other factors [3]. No single algorithm is best in all scenarios.

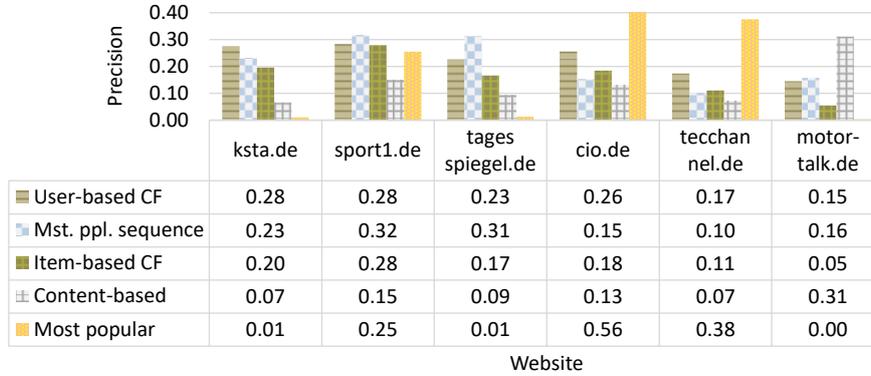

**Fig. 1.** The precision of five recommendation algorithms on six news platforms. On almost every platform, a different algorithm performs best [3]

To identify the most effective recommendation algorithm for a given use-case, practitioners typically assess a pool of algorithms for suitability. These algorithms are trained using historical data, and their effectiveness is estimated using cross-validation techniques. The best candidate-algorithm is typically chosen for a scenario on the basis of having the lowest predictive error, or highest accuracy [16]. Single-number metrics such as precision, recall, nDCG, RMSE, and click-through rate, are used.

We argue that focusing on the overall single-number performances of algorithms is not optimal. If we were able to accurately choose the most effective recommendation algorithm for a given user, item, and context, the overall effectiveness of the recommender system should improve.

Meta-learning for algorithm selection aims to predict the best algorithm to use in a given scenario. It does this by learning the relationship between characteristics of data, and the performance of algorithms for that data [18]. It is useful in situations where distinct algorithms perform differently in varied scenarios.

Meta-learners for algorithm selection can be trained and used at different levels of analysis. We develop here the following distinctions between these meta-learners:

1. *Global-level* meta-learners use the characteristics of data to select the overall-best algorithm for entire datasets, or an entire platform (e.g. a particular news website)
2. *Mid-level* meta-learners select the overall-best algorithm for subsets of data. For example, users in a recommendation scenario may differ by how many explicit ratings they have made. Collaborative filtering algorithms are inappropriate for users who have made no ratings. Mid-level meta-learners may operate on any such entity (users, items, gender, age, etc.). They may also operate on arbitrary subsets of data (e.g. clusters selected via unsupervised learning).
3. *Micro-level* meta-learners attempt to select the best algorithm for every instance in a dataset, or every single recommendation request on a given platform.

In this paper, we propose an application of micro-level meta-learning to recommendation [2]. Most existing meta-learning approaches for recommendation act at a global-level; they attempt to select the single-best algorithm for entire datasets. We attempt to select the best algorithm separately for each dataset instance.

## Related Work

Meta-learning has been used for algorithm-selection in recommender systems [1, 7–9, 12, 19]. These authors manually define meta-features, which aggregate information from datasets into single-number statistics. For example, the number of instances in the dataset is a 'simple' meta-feature, the mean or kurtosis of a column is a 'statistical' meta-feature. They then use supervised machine learning to learn the relationships between the meta-features and the performance of recommendation algorithms on datasets, measured by standard metrics. Most authors attempt to select the single-best algorithm for entire datasets; they are proposing global-level meta-learners. This is beneficial; it simplifies the process of choosing an algorithm for a recommendation scenario, for example. However, it results in non-optimal recommender system performance, as the best algorithm for each instance is not selected.

Ekstrand and Riedl [10] propose a mid-level meta-learner; they attempt to select the best algorithm for subsets of data in a dataset. They construct an ensemble of a small number of diverse algorithms and compare the ensemble's performance to baseline algorithms. Using a classifier and one meta-feature, an attempt is made to choose the best algorithm between item-item and user-user collaborative filtering for each user. Their meta-learning approach did not outperform the overall-best algorithm (RMSE; meta-learner ~0.78, item-item CF: ~0.74).

Ensemble approaches exist that uniquely combine several algorithms for each instance, such as stacked generalization [22], and feature-weighted linear stacking [20]. These methods have produced good results [13, 17, 21], however they require that all algorithms in the pool are executed before their output is combined, or before a single algorithm is selected. For a large pool of algorithms, this may be a prohibitive requirement.

To the best of our knowledge, there are no applications of meta-learning for recommender systems that select algorithms at a more granular level than per-user.

## Methodology

Our meta-learner aims to use the best algorithm for each user-item pair in a recommendation dataset, from a pool of single recommendation algorithms. The pool of algorithms used in our system includes nine collaborative filtering algorithms from the Surprise recommendation library[2]:

1. Co-clustering
2. KNN (Baseline)
3. KNN (Basic)
4. KNN (with Means)
5. Non-negative Matrix Factorization (NMF)
6. SVD
7. SVD++
8. Slope One
9. Baseline - A collaborative filtering baseline which always predicts the overall-average rating, biased by the overall-average rating for the user, and overall-average rating for the item

We performed our experiments using the MovieLens 100K and 1M datasets [14]. We chose these datasets over others as they contain demographic information for users, and detailed item information, which may be useful in attempting to select a recommendation algorithm (**Table 1**).

To assess the potential improvements that a hypothetically 'perfect' micro-level meta-learner could offer over an overall-best algorithm, we first evaluate our pool of algorithms individually on the MovieLens 100K and 1M datasets. We randomly divide each dataset with a 70%:30% training:test split. We train each algorithm, and for each user-item pair in the test set we note the error between the predicted rating and true rating. We calculate the overall RMSE for each algorithm. We also note which algorithm performed best for each user-item pair; a perfect micro-level meta-learner would be able to choose this algorithm. This process is illustrated in Table 1. We calculate the overall RMSE that a perfect micro-level meta-learner would achieve.

In a second evaluation, we test our micro-level meta-learner. We randomly divide each dataset into two equal subsets: a training set and an evaluation set. The training set is used to train our nine individual collaborative filtering recommendation algorithms, resulting in nine ready-to-use models. To evaluate the meta-learner, we perform a 5-fold cross validation on the evaluation set. Each fold splits the evaluation set into two subsets: a meta-training set, and a test set. The trained models of the individual recommendation algorithms are applied to the meta-training set, resulting in errors of the algorithms. The meta-learner is then trained on these errors, resulting in the meta-model. Finally, the meta-model is tested on the test set. We further assess a 'perfect' meta-learner using this evaluation set. We use RMSE to evaluate our approach.

---

[2] http://surpriselib.com/

**Table 1:** An illustration of the MovieLens dataset. The algorithm with the best predicted error rate per row is highlighted. A hypothetical perfect meta-learner should predict this algorithm.

| ID | User attributes ||||||| Movie attributes ||||||||| Rating | Predicted Ratings (Error Rates) ||||
|---|---|---|---|---|---|---|---|---|---|---|---|---|---|---|---|---|---|---|---|---|---|---|
| | User ID | Age | Gender || Occupation |||| Movie ID | Release Year ||||  Genre |||| | SVD | Slope One | KNN Basic | ... |
| | | | Male | Female | artist | doctor | educator | ... | | 60s | 70s | 80s | ... | Action | Comedy | Crime | ... | | | | | |
| 1 | 506 | 46 | 1 | 0 | 1 | 0 | 0 | ... | 568 | 0 | 0 | 0 | ... | 1 | 0 | 0 | ... | 5 | -1.06 | -1.08 | -1.12 | ... |
| 2 | 363 | 20 | 1 | 0 | 1 | 0 | 0 | ... | 849 | 0 | 0 | 0 | ... | 1 | 0 | 0 | ... | 2 | 0.26 | 0.23 | 0.66 | ... |
| 3 | 842 | 40 | 1 | 0 | 0 | 1 | 0 | ... | 874 | 0 | 0 | 0 | ... | 0 | 0 | 0 | ... | 5 | -1.94 | -2.09 | -2.20 | ... |
| 4 | 312 | 48 | 1 | 0 | 0 | 0 | 0 | ... | 241 | 0 | 0 | 0 | ... | 1 | 0 | 0 | ... | 3 | 1.09 | 0.87 | 0.89 | ... |
| 5 | 42 | 30 | 1 | 0 | 0 | 0 | 1 | ... | 294 | 0 | 0 | 0 | ... | 0 | 1 | 0 | ... | 4 | -0.68 | -0.82 | -0.51 | ... |
| 6 | 812 | 22 | 1 | 0 | 0 | 0 | 0 | ... | 326 | 0 | 0 | 0 | ... | 1 | 0 | 0 | ... | 4 | -0.72 | -0.33 | -0.51 | ... |
| 7 | 450 | 35 | 0 | 1 | 0 | 0 | 1 | ... | 3 | 0 | 0 | 0 | ... | 0 | 0 | 0 | ... | 4 | -0.37 | -0.54 | -0.87 | ... |
| 8 | 90 | 60 | 1 | 0 | 0 | 0 | 1 | ... | 42 | 0 | 0 | 0 | ... | 0 | 1 | 0 | ... | 4 | 0.17 | 0.10 | -0.03 | ... |
| 9 | 36 | 19 | 0 | 1 | 0 | 0 | 0 | ... | 882 | 0 | 0 | 0 | ... | 0 | 0 | 0 | ... | 5 | -1.24 | -0.71 | -1.43 | ... |
| 10 | 551 | 25 | 1 | 0 | 0 | 0 | 0 | ... | 235 | 0 | 0 | 0 | ... | 1 | 1 | 0 | ... | 1 | 1.97 | 2.16 | 1.79 | ... |
| 11 | 553 | 58 | 1 | 0 | 0 | 0 | 1 | ... | 1126 | 0 | 0 | 0 | ... | 0 | 0 | 0 | ... | 4 | -0.53 | -0.61 | -0.95 | ... |
| 12 | 168 | 48 | 1 | 0 | 0 | 0 | 0 | ... | 1028 | 0 | 0 | 0 | ... | 0 | 1 | 0 | ... | 2 | 0.80 | 0.88 | 1.34 | ... |

During training, we perform the following steps (**Fig. 2**):

1. Single algorithms are trained on the training set. The result is a single model for each collaborative filtering algorithm. This model can predict a rating for a given user-item pair.
2. Each trained single algorithm is applied to every row in the meta-learner training set. This gives us a rating-prediction for each algorithm, for training set rows.
3. For each single algorithm and each row in the meta-learner training set, a rating-prediction error is calculated. This error is the difference between the predicted rating and true rating. These rating-prediction errors are illustrated in Table 1.
4. Using the meta-learner training set, we train a linear model for each single algorithm. These linear models are trained on the content-features from rows, for example: the gender, age and occupation of the user, genre and year of the movie. We also include 10 meta-features: the rating mean, standard deviation, minimum, maximum and median, for each user and each item. Categorical features are one-hot-encoded. The training target is the rating-prediction error from Step 3. These models allow us to predict each algorithm's error, based on the content-features and meta-features of the user-item pair.

In the prediction stage of our meta-learner, the rating for a given user-item pair is calculated in the following steps:

1. For each algorithm in the pool we predict the error that the algorithm will make for this user-item.
2. We rank the algorithms according to the absolute value of their predicted errors. We choose the algorithm with the lowest predicted error.
3. The chosen algorithm is applied to the user-item pair. The rating predicted by the chosen algorithm is the final output of our system.

We compare our meta-learner to two baselines. The first is the best single algorithm for the dataset. The second is a simple ensemble, which averages the ratings predicted by all single algorithms for each row.

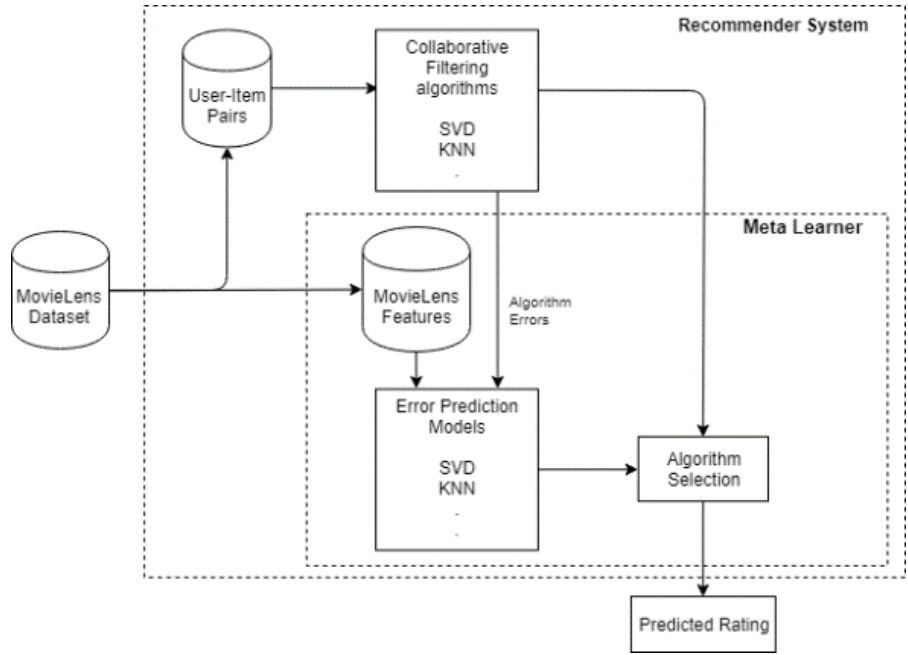

**Fig. 2.** A block diagram of our Meta-learned Recommender System.

## Results & Discussion

The results from our first evaluation of a hypothetical 'perfect' micro-level meta-learner are shown in **Fig. 3** and **Fig. 4**.

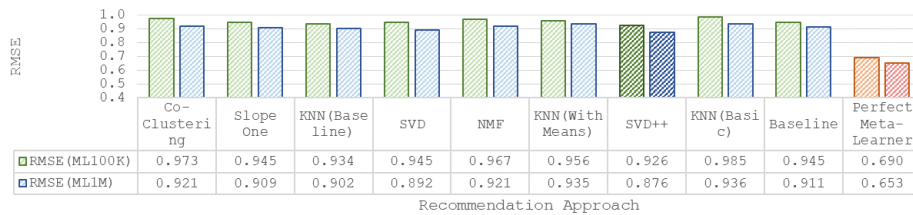

| Recommendation Approach | Co-Clustering | Slope One | KNN (Baseline) | SVD | NMF | KNN (With Means) | SVD++ | KNN (Basic) | Baseline | Perfect Meta-Learner |
|---|---|---|---|---|---|---|---|---|---|---|
| RMSE (ML100K) | 0.973 | 0.945 | 0.934 | 0.945 | 0.967 | 0.956 | 0.926 | 0.985 | 0.945 | 0.690 |
| RMSE (ML1M) | 0.921 | 0.909 | 0.902 | 0.892 | 0.921 | 0.935 | 0.876 | 0.936 | 0.911 | 0.653 |

**Fig. 3.** The average error (RMSE) of different collaborative-filtering algorithms on the MovieLens 100K and 1M datasets. The RMSE for a 'perfect' micro-level meta-learner, in which the best algorithm is chosen for each user-item pair, is shown.

For both MovieLens 100K and 1M, the algorithms with the lowest RMSE from our collection of collaborative filtering algorithms are SVD++ (RMSE; ML100K: 0.926, ML1M: 0.876), followed by a variant of k-nearest neighbors (KNN Baseline) (RMSE; ML100K: 0.934) and SVD (ML1M: 0.892) (**Fig. 3**).

An evaluation of MovieLens test-sets using these algorithms would suggest to an operator that SVD++ and KNN (Baseline) are the best candidate algorithm to use. However, for each row in the 100K dataset, SVD++ is not the best algorithm most often (ML100K; SVD++: 15.85%, vs. KNN (Basic): 16.7%) (**Fig. 4**). The second-best algorithm by RMSE (KNN (Baseline)) is least often the best algorithm for each user-item in the 100K dataset. In the 1M dataset, the second most-frequently accurate algorithm KNN (Basic) (16.10%), is the least accurate with regards to RMSE (0.936).

Using the overall-best algorithms for these datasets is therefore a significant compromise. In a hypothetical scenario in which the best algorithm per dataset instance could be chosen – i.e. with a perfect meta-learner – RMSE would be improved by 25.5% for both 100K and 1M when compared to their respective overall-best algorithms (**Fig. 3**).

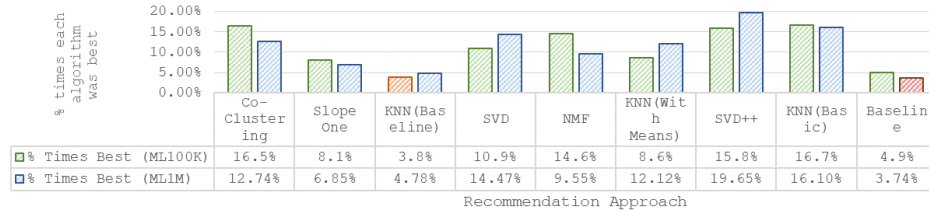

**Fig. 4.** Percentage of times each collaborative filtering algorithm was best on the MovieLens 100K and 1M datasets. The algorithms least often best are highlighted (ML100K: KNN (Baseline); ML1M: Baseline).

The results of our second evaluation are presented in **Fig. 5**. Our meta-learner (RMSE, 100K: 0.973; 1M: 0.908) performed 2-3% worse than the best individual algorithm SVD++ (RMSE, 100K: 0.942; 1M: 0.887) and the average-rating baseline (RMSE, 100K: 0.943; 1M: 0.893). These results suggest that the current implementation of our meta-learner is unable to accurately rank algorithms according to their rating errors.

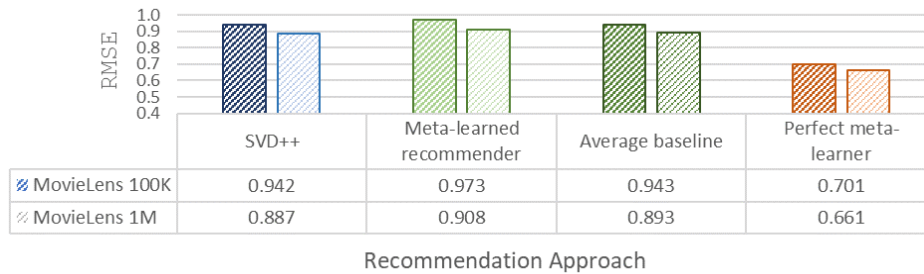

**Fig. 5.** Effectiveness of the meta-learned recommender system, compared to 1) the overall-best algorithm, 2) an ensemble which predicts an average rating of all algorithms per row, 3) the theoretical best case in which the actual best algorithm per user-item pair

The nine algorithms used are of a similar class. We expect that a more heterogenous pool of algorithms may provide better results. It is also possible that, because the algorithms we used have been trained on the same dataset that our meta-learner has been trained on, they are all already well fitted to the data. In such a case, the differences between error-predictions and real errors might be too small to allow for accurate rankings of the top algorithms: our predictions for the worst algorithm per row are twice as good as our predictions for the best (Accuracy; 0.12 vs 0.21). Linear regression may also not be suitable to model rating-prediction errors. More advanced algorithms may be more suitable for meta-learning.

Our approach is computationally inexpensive compared to standard ensembles. As the final prediction is calculated by one chosen algorithm, we do not need to obtain and weight predictions from all algorithms in a pool to make our final prediction. We also do not need to use all algorithms to retrain our system when a new algorithm is introduced, as the rating-prediction error models are independent of each other. For these reasons, in future work we hope to improve our rating-prediction error models.